\RequirePackage{fix-cm}
\documentclass[twocolumn,epjc3]{svjour3_arxiv}
\usepackage[symbol]{footmisc}

\smartqed  
\RequirePackage{graphicx}
\RequirePackage[caption=false]{subfig}
\RequirePackage{xcolor}
\usepackage{bm}
\usepackage{xspace}
\usepackage{multirow}
\usepackage{xcolor}
\usepackage[british]{babel}
\usepackage{hyphenat}
\usepackage{amsmath,amssymb}

\usepackage{graphicx} 
\usepackage{caption}
\usepackage{subfig}
\usepackage{float} 
\graphicspath{{figures/}} 
\usepackage{booktabs}
\usepackage{hyperref}

\usepackage{eurosym}

\newcommand{\beq}{\begin{equation}}
\newcommand{\eeq}{\end{equation}}

\def\qgs{\textsc{QGSJet}-II.04\xspace}

\journalname{Eur. Phys. J. C}
\begin{document}

\title{The gamma/hadron discriminator $LCm$ in realistic air shower array experiments
}


\author{
        R. Concei\c{c}\~ao\thanksref{LIP,IST,e1}
        \and
        P. J. Costa\thanksref{LIP,IST}
        \and
        L. Gibilisco\thanksref{LIP,IST}
        \and 
        M. Pimenta\thanksref{LIP,IST} 
        \and
        B. Tom\'e\thanksref{LIP,IST}
}

\thankstext{e1}{e-mail: ruben@lip.pt}


\institute{Laboratório de Instrumentação e Física Experimental de Partículas (LIP), Lisbon, Portugal \label{LIP}
\and
 Instituto Superior T\'{e}cnico (IST), Universidade de Lisboa, Lisbon, Portugal \label{IST}
}

\date{Received: date / Accepted: date}

\maketitle

\begin{abstract}

In this article, it is shown that the $C_k$ and $LCm$ variables, recently introduced as an effective way to discriminate gamma and proton-induced showers in large wide-field gamma-ray observatories, can be generalised to be used in arrays of different detectors and variable fill factors. In particular, the $C_k$ profile discrimination capabilities are evaluated for scintillator and water Cherenkov detector arrays.

\end{abstract}

\section{Introduction}
\label{sec:intro}

At energies surpassing approximately $100\,$GeV, gamma-rays originating from astrophysical sources cannot be directly detected. Instead, their detection is inferred indirectly by reconstructing the extensive air showers (EAS) they generate upon interacting with Earth's atmosphere~\cite{VHE_gamma_rays}. A careful evaluation of the shower characteristics is essential to differentiate EAS induced by gamma rays from those produced by the dominant cosmic ray background. A noteworthy parameter in this discrimination process is the assessment of a shower's muon content, which is expected to be higher for showers triggered by hadronic processes than gamma rays.

Although muon counting offers a highly effective method for discriminating between gamma and hadron-induced showers (see for instance~\cite{LHAASO}, its practical implementation necessitates some form of shielding – such as a layer of soil or water above the detector – to absorb the electromagnetic component of the shower. This requirement renders such experiments financially demanding and largely unfeasible in ecologically sensitive regions.

A recent study, documented in \cite{LCm}, demonstrated that crucial information for discriminating between gamma and hadron-induced showers can be derived from the azimuthal asymmetry of the ground-level shower footprint. Through comprehensive simulation investigations, the discriminatory potential of the newly introduced observables, denoted as $LCm$, is comparable to that achieved through muon counting. More importantly, these observables offer significantly improved experimental accessibility, addressing the challenges posed by the need for shielding in muon-based methods. 

According to~\cite{LCm}, the assessment of the fluctuations is done through the quantity 
$C_{k}$, defined in circular rings $k$ centred around the shower core position, with a width of $10$\,m and a mean radius $r_{k}$ as:

\begin{equation}
C_{k} =\frac{2}{n_{k}(n_{k}-1)} 
\frac{1}{\left<S_{k}\right>}\sum_{i=1}^{n_{k}-1}\sum_{j=i+1}^{n_{k}}(S_{ik}-S_{jk})^{2} ,   
\label{eq:CK}
\end{equation}

where $n_{k}$ is the number of stations in the ring $k$, $\left<S_{k}\right>$ is the mean signal in the stations of the ring $k$, and $S_{ik}$ and $S_{jk}$ are the collected signals in the stations $i$ and $j$ of the ring $k$, respectively.

The shower azimuthal asymmetry level is stated by the quantity $LCm$ defined  as the value of a  parametrisation  of the $\log(C_{k}) $ distribution at a given value of $ r_{k} = r_{m} $, and $r_{m}$ was fixed to $r_{m} = 360\,$m.

The behaviour of $LCm$ has been studied in~\cite{LCm} as a function of a scaling factor defined as such:

\begin{equation}
    K = E^{\beta} \times  FF  ,
 \label{eq:ruleThumb}
\end{equation}

where $E$ is the primary  energy (in TeV), $\beta$  is the index of the power dependence of the mean number of muons at the ground and $FF$ is the fill factor, the fraction of instrumented array area. The parameter $\beta$ was fixed to $0.925$, a typical mean value used in hadronic shower simulations. It has been shown that, for different energies and fill factors,
but identical $K$ factors, the $C_k$ distributions are essentially the same. 

In~\cite{LCm}, a uniform array of single-layer water Cherenkov detectors (WCDs) stations with a constant fill factor  and each with four photomultiplier tubes (PMTs) at its bottom was considered; an overview of these results will be given in section \ref{sec:lcmWCDs}. 

Motivated by studies that try to use $C_k$ and $LCm$ in realistic experiments, such as its application to KASCADE data~\cite{KASCADE_LCm}, this study has been hereby extended to setups with similar single-layer WCDs with the same radius and height, but with different numbers of photo-sensors at their bottom (section \ref{sec:vem}).  The applicability of the $C_k$ variable in scintillator arrays has been assessed as well in sections \ref{sec:scint} and \ref{sec:scintPb}.
Finally, a simple way to handle arrays with variable $FF$ is discussed in section \ref{sec:variableFF}.

The results presented in this work were obtained using air shower simulations whose output was subsequently processed to emulate the behaviour of a detector array.
CORSIKA (version 7.5600) \cite{CORSIKA} was used to simulate gamma-ray and proton-induced vertical showers assuming an observatory altitude of $5200\,$m a.s.l., and FLUKA~\cite{fluka,fluka2} and \qgs \cite{qgs} were used as hadronic interaction models for low and high energy interactions, respectively. The current investigations were conducted within a specific gamma-ray energy range, encompassing a single energy bin of width $\log(E) = 0.2$, originating at $100\,$TeV.
The energies of the generated proton samples were chosen so that the total electromagnetic signal at the ground would be similar to the gamma-ray-generated showers.

A 2D-histogram emulated the ground detector array with cells with an area of $\sim 12\,{\rm m^2}$ covering the available ground surface. Each cell represents one station. The signal in each station was estimated as the sum of the expected signals due to the particles hitting the station, using calibration curves for each station type as a function of the particle energy for protons, muons and electrons/gammas. Fluctuations induced by the detector were mimicked by applying Gaussian distributions centred on the values given by the calibration curves and with sigmas equal to the respective RMS.
The fill factor of the array was set in the interval $\in ]0,1]$, by masking the 2D-histogram with the appropriate regular pattern. Following reference~\cite{LCm}, it was chosen for the $\sim 100\,$TeV simulation a sparse array with a fill factor of $12\%$.

The same methodology as previously outlined will be employed to analyze the scintillator arrays, both with and without the inclusion of a lead converter.

\section{$LCm$ and the detector technology}
\label{sec:lcmDetector}

\subsection{Water Cherenkov detectors}
\label{sec:lcmWCDs}
\begin{figure}[!t]
  \centering
  \includegraphics[width=0.45\textwidth]{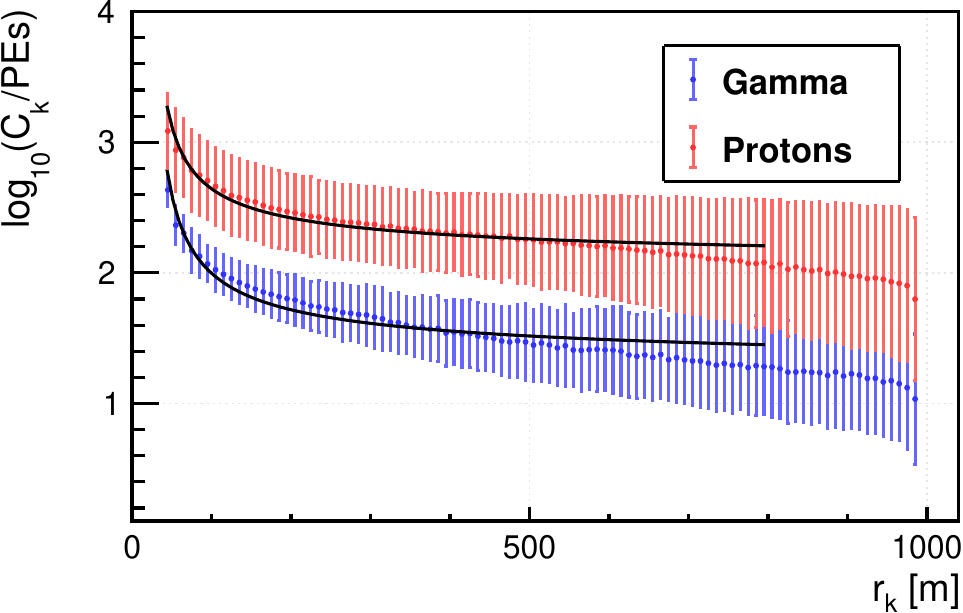}
  \caption{\label{fig:ck3pmt} $C_k$ profile as a function of the distance to the shower core computed for showers with energies of $\sim 100\,{\rm TeV}$, using an array of Mercedes WCD stations with $FF=12\%$.}
\end{figure}

\begin{figure}[!t]
  \centering
  \includegraphics[width=0.45\textwidth]{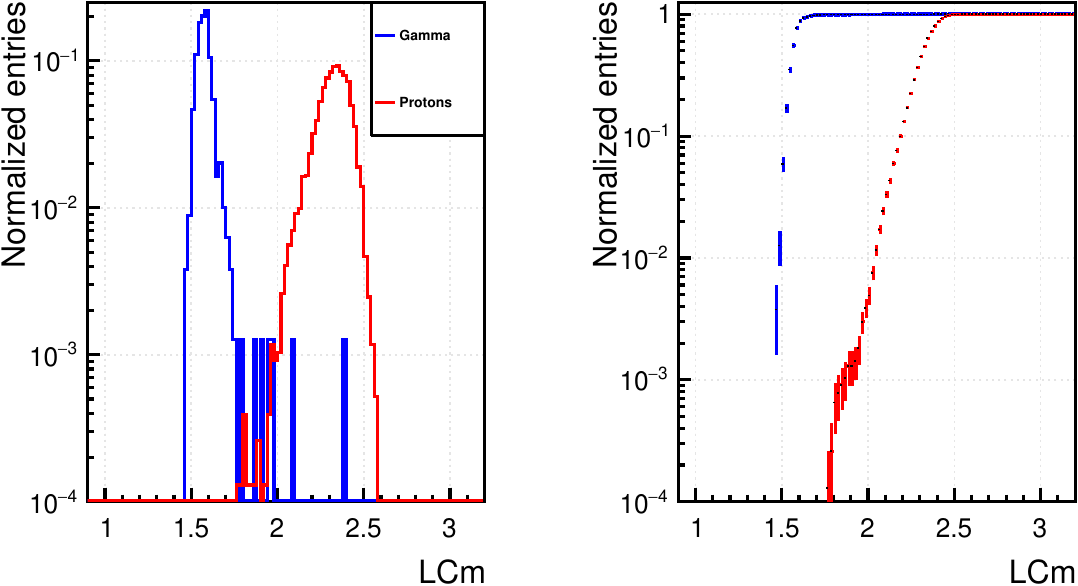}
  \caption{\label{fig:lcm3pmt} $LCm$ distribution (left) and cumulative (right) for showers with energies of $\sim 100\,{\rm TeV}$, using an array of Mercedes WCD stations with $FF=12\%$.
}
\end{figure}

Due to the presence of hadronic sub-showers in hadron-induced extensive air showers, the $C_k$ variable, as defined in equation \ref{eq:CK}, is expected to be larger for proton-induced showers compared to gamma-induced showers with equivalent energies at the ground. This has been verified in \cite{LCm}, where the $C_k$ variable has been computed for proton and gamma showers using a uniform array of WCDs, each equipped with 4 PMTs at the bottom. The same results have been hereby replicated with an array of Mercedes (3 PMTs at the bottom) WCDs \cite{Mercedes} of the same shape and size and a uniform $FF$ of 12\% for gamma and proton showers with primary energy around $100\,{\rm TeV}$. 
The detailed analysis of the dependence of $C_k$ from the number of PMTs in the WCDs is presented in section~\ref{sec:vem}. 
In Figure~\ref{fig:ck3pmt}, it is shown the distributions of the mean values of  $\log (C_{k})$  are represented as a function of the radius $r_{k}$ for both primaries. The error bars depict the standard deviation of the distributions.
The $LCm$ has then been computed, and it is shown in Figure~\ref{fig:lcm3pmt}. At an efficiency close to $100\%$, next to no background events are left within the current limits of the statistics\footnote{While the present study was done with $\mathcal{O}(10^4)$ events for proton showers, in a recent study, just submitted for publication~\cite{LCmTail}, it is demonstrated using a simulation set with $\mathcal{O}(10^6)$ event that $LCm$ has a discrimination capability slightly higher that the measurement of the EAS muon content.}.

\subsection{Scintillators}
\label{sec:scint}
Scintillator arrays are widely used in cosmic-ray observatories. Without entering a detailed discussion of their advantages or disadvantages as compared to WCDs arrays, 
hereafter, their performance in gamma/hadron discrimination is explored regarding the gamma/hadron discriminating variables, $C_{k}$ and $LCm$.
 
Scintillators are very good for tagging charged particles but not for measuring their energies and/or identifying muons without using shielding to the other charged particles. Therefore gamma/hadron discriminating algorithms based on muon counting are not efficient for unshielded scintillator arrays. Furthermore, scintillators are mostly insensitive to the shower secondary photons.

However, as discussed in reference \cite{LCm}, the strong correlation between $LCm$ and the total number of muons hitting the detectors is still present without considering the contribution to the signal from the muons. Therefore, in this section, we investigate the possibility of using $LCm$, measured in scintillator arrays, as a gamma/hadron discriminator.

 In the present simulation framework (see section \ref{sec:intro}), the emulation of scintillator arrays can be easily done by introducing new calibration curves corresponding to their response to protons, muons, electrons and gammas as a function of the particle energy. 

To be as realistic as possible, these calibration curves were built from the mean expected
signal of a simulation of the plastic scintillator detectors using the Geant4 toolkit
\cite{GEANT4:2003,GEANT4:2006,GEANT4:2016}, in particular its capabilities to simulate the optical processes and describe the optical properties of the materials.
The scintillator is $1\,$cm thick and $50\,$cm long. The light is readout at both ends by a photodetector with the same sensitive area as the Hamamatsu R9420, with a 38 mm bialkali photocathode. All relevant optical parameters  \cite{AugerSSD} were implemented in this simulation, namely the scintillator's light yield, the emission spectra and the
quantum efficiency of the PMT's photocathode. The optical properties of a white diffuser, wrapping the scintillator, were also included, using the unified model \cite{Unified} implemented in the Geant4 toolkit.

\begin{figure}[!t]
  \centering
  \includegraphics[width=0.45\textwidth]{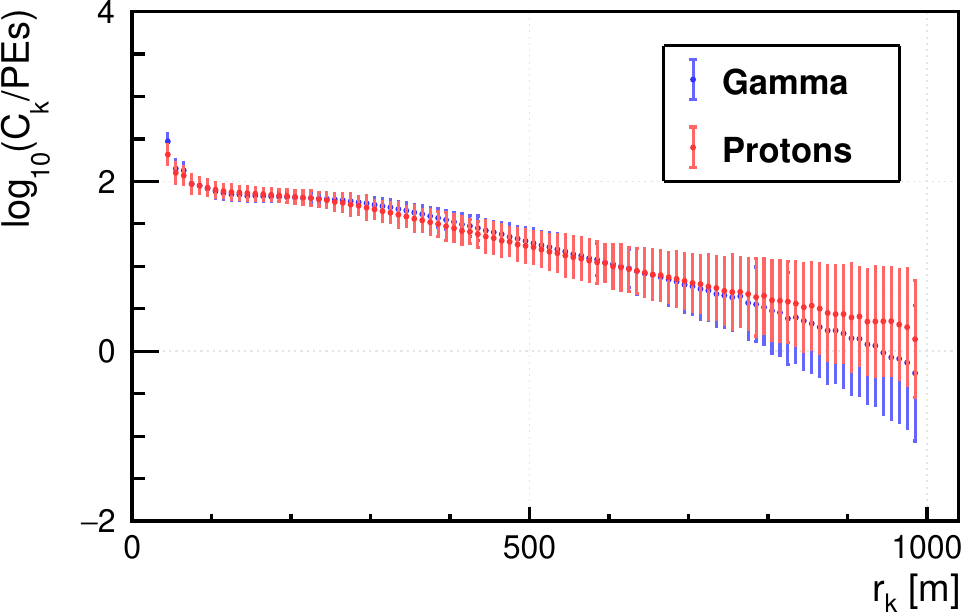}
  \caption{\label{fig:scint} $C_k$ distributions for showers with energies of $\sim 100\, {\rm TeV}$, using an array of scintillators with $FF=12\%$.
}
\end{figure}

Figure~\ref{fig:scint} shows the $\log (C_k)$ distributions as a function of the radius $r_{k}$ for $\sim100\,{\rm TeV}$ gamma and proton-induced showers with similar energy at the ground, considering a scintillator array with a $FF=12\%$. From this figure, it can be seen that the $C_k$ profile does not depend on the nature of the primary particle and, therefore, cannot be used as a gamma/hadron discriminator.

\subsection{Scintillators coupled to lead converters}
\label{sec:scintPb}
\begin{figure}[!t]
  \centering
  \includegraphics[width=0.45\textwidth]{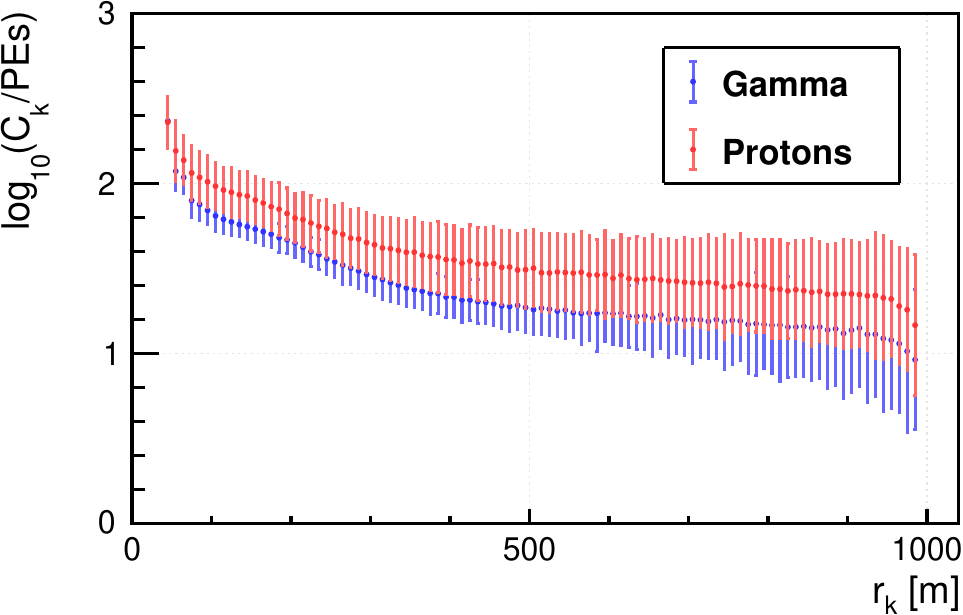}
  \caption{\label{fig:scintpb} $C_k$ distributions for showers with energies of $\sim 100\, {\rm TeV}$, using an array of lead-shielded scintillators with $FF=12\%$.
}
\end{figure}

As a further exercise, to enhance the scintillator response to  shower photons,  a thin, $1\, \mathrm{X_0}$ thick lead layer was placed on the top of the scintillators.
The situation improves compared to the unshielded scintillators. In this case, the number of electrons produced during the ionization losses in the lead plate will scale with energy, making the apparatus relatively sensitive to the shower calorimetry. Consequently, $C_k$ will differ for gamma and hadron-induced showers, as seen in Figure~\ref{fig:scintpb}, thus enabling gamma/hadron discrimination. However, it should be noted that the WCD has a stronger discrimination power, as can be seen by comparing the separation between primaries in Figures~\ref{fig:ck3pmt} and \ref{fig:scintpb}.

\section{Impact on the number of photosensors}
\label{sec:vem}
Having established the WCDs as a better choice than scintillators in terms of gamma/hadron discrimination capabilities, the study of the $C_k$ variable is now extended to WCDs with different numbers of photosensors. The PMTs are all placed at the bottom of the tank. The dimensions of the station are the same for all tested PMT configurations: a radius of  $2 {\rm m}$ and a height of $1.7\,{\rm m}$. In particular, arrays of 4-PMTs and Mercedes stations and arrays of stations with a single central PMT at their bottom (hereafter designated as Mercedes-1) are studied. 

To simplify the comparison of  results obtained using setups with stations with a different number of PMTs, $C_{k}$ and $LCm$ variables are normalised, in each setup, by the corresponding mean signal produced by one relativistic vertical muon crossing one  station at its centre (VEM), $Q_{\rm VEM}$, 
and renamed as $C_{k}^\ast$ and $LCm^\ast$:

\begin{equation}
C_{k}^\ast =\frac{C_{k}}{Q_{{\rm VEM}}},   
\label{eq:CK*}
\end{equation}

\begin{equation}
LCm^\ast =\frac{LCm}{Q_{\rm VEM}}.   
\label{eq:LCM*}
\end{equation}

\begin{figure}[!t]
  \centering
  \includegraphics[width=0.45\textwidth]{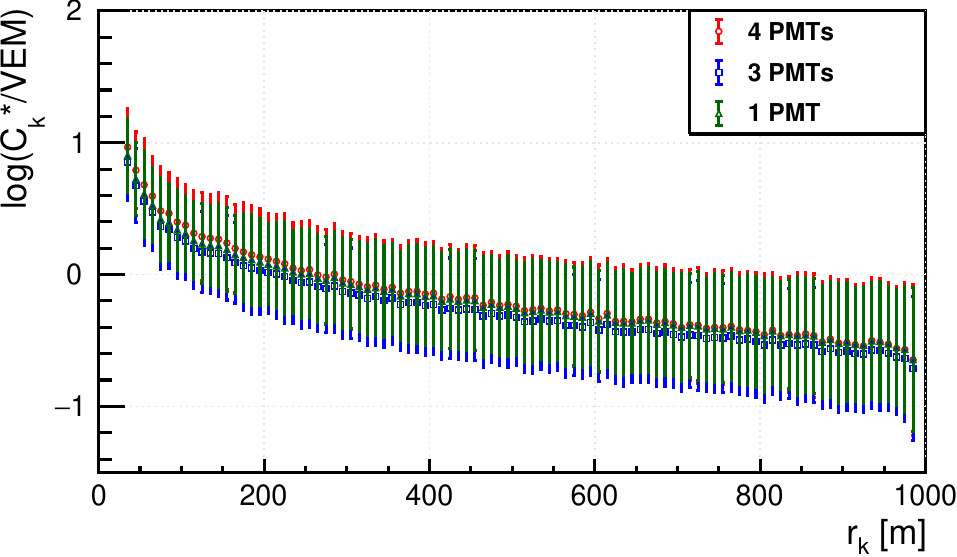}
  \caption{\label{fig:vemCk} $C_k^*$ distributions for proton showers with energies of $\sim 100\, {\rm TeV}$ using a $FF = 12\%$ array of WCD stations equipped with four PMTs (red circles), three PMTs (blue squares) and one PMT (green triangles).
}
\end{figure}

\begin{figure}[!t]
  \centering
  \includegraphics[width=0.45\textwidth]{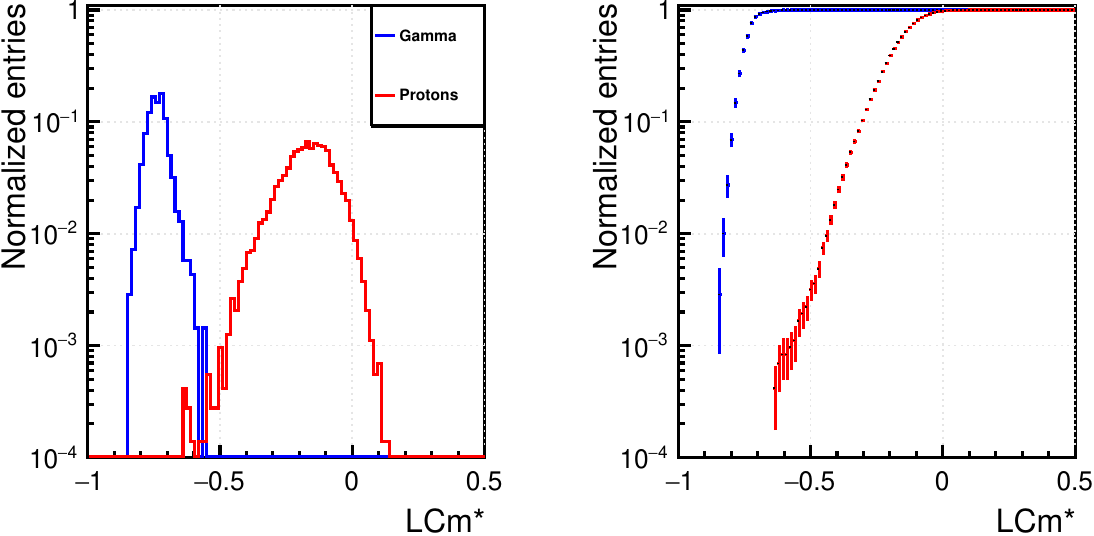}
  \caption{\label{fig:lcm1pmt} $LCm$ distribution (left) and cumulative (right) for showers with energies of $\sim 100 {\rm TeV}$, using an array of Mercedes-1 WCD stations with $FF=12\%$.
}
\end{figure}

It should be noted that each station (with a different number of PMTs) will have a specific $Q_{VEM}$. This value can be obtained using dedicated measurement~\cite{AugerRPC} or the analysis of the omnidirectional atmospheric muons~\cite{AugerVEM}.
According to its definition, the mean of $C_{k}^\ast$ should not depend too much on the number of the PMTs placed at the bottom of each station as long as the expected mean signal of the station is high enough not to introduce significant statistical fluctuations.
Indeed, this behaviour is confirmed in Figure~\ref{fig:vemCk}, where it is shown, for proton showers, the  $\log(C_{k}^*)$ distributions for identical WCD stations but different numbers of PMTs.
After the VEM normalisation, the differences become quite small.
From the above considerations, an array of Mercedes-1 WCDs with a water height of 1.7m  would guarantee the required level of gamma/hadron discrimination (rejection factors of the order or higher than $10^{-4}$) for energies and $FF$ ensuring a scaling factor $K$ (eq. \ref{eq:ruleThumb}) of about 5 to 10.
Such a premise is verified in Figure~\ref{fig:lcm1pmt}, where the  $LCm^\ast$ distributions, as well as their cumulative distributions, are shown for gamma  (blue points) and proton-induced showers (red points). The primary energies of gamma showers are $\sim 100{\rm TeV}$, and the proton showers have been selected to have a similar energy at the ground, while the array of Mercedes-1 stations has $FF=12\%$.

\section{$LCm$ computation in inhomogeneous arrays}
\label{sec:variableFF}

To cover a wide energy range, the layout of many present and future gamma-ray Observatories has a high $FF$ in the inner regions (the so-called compact arrays), primarily intended to cover the lower energy region, and a low $FF$ in the outer regions (usually designated as sparse arrays), conceived mainly to reach the higher energies. Ideally, the transition between these two regions should be smooth to optimise the observatory's sensitivity to intermediate energies.

\begin{figure}[!t]
  \centering
  \includegraphics[width=0.45\textwidth]{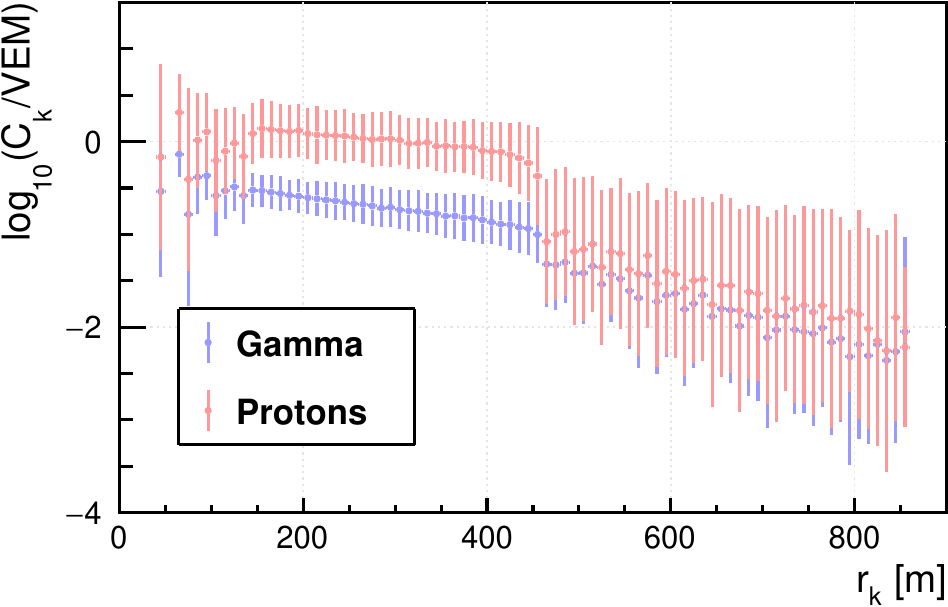}
  \caption{\label{fig:ckFFnocorr} $C_k^*$ distributions for showers with energies of $\sim 100\,{\rm TeV}$ using an array of Mercedes WCD stations centred $300{\rm m}$ away from the shower core and composed of a dense ($FF=100\%$) $160\,{\rm m}$-wide central region surrounded by a $560\,{\rm m}$-wide ring with $FF = 12\%$. 
}
\end{figure}

In all these layout designs, the fill factor will not be constant throughout the array, inducing discontinuities in the $C_{k}$ distributions, not present beforehand. This effect is particularly evident in Figure~\ref{fig:ckFFnocorr}, where the shower core was placed at a distance of $300\,$ meters from the centre of the array. This particular array is composed of two fill factors: a dense array ($FF = 100\%$) with a radius of $160\,$ meters, encircled by a sparse array with a radius of $560\,$ meters and a $FF$ of $12\%$. It is important to note that in addition to the observed discontinuities, the error bars in this figure are notably larger compared to those presented in Figure~\ref{fig:ck3pmt}. This discrepancy arises due to the increased number of stations involved in the computation of $C_k$ in the latter case.

To handle these discontinuities, an effective fill factor in each ring is introduced. This factor is defined as:

\begin{equation}
 FF_k =  \frac{n_k} {n_{k_1}} ,
\label{eq:FF_k}
\end{equation}

where $n_{k}$ is, as before, the number of stations in ring $k$, while $n_{k_1}$ is the number of stations in the ring $k$ if the $FF$ is $100\%$.

Consequently, equation \ref{eq:ruleThumb} gets redefined for each ring as:

\begin{equation}
 K_k = E^{\beta} \times  FF_k  .
 \label{eq:ruleThumb_k}
\end{equation}

According to equation 4.3 of~\cite{LCm}, $LCm$ is a function of $K$ and may be parameterised for the proton sample as:

\begin{equation}
f_p(K)= LCm_p (K) \sim  A_p  + \frac{B_p}{\sqrt{K}} 
\label{eq:LCm*_K}
\end{equation}

where $A_p$ and $B_p$ are constants defined for the primary proton sample.

The function $f_p(K)$ can now be used as a correction factor: 

\begin{equation}
C_{kcor} = C_{k} 10^{(f_p(K_{ref})-f_p(K_k))},
 \label{eq:Ck*_cor}
\end{equation}

where $K_{\rm ref}$ and $K_k$ are computed using equation~\ref{eq:ruleThumb_k}, but using for the $FF_k$, respectively, a reference value (typically the mean $FF$ of the array) and the effective $FF_k$ of that specific ring $k$ (equation \ref{eq:FF_k}). 
Such a correction factor has to be applied whenever the effective $FF_k$ is not the reference $FF$, as in the case where two or more regions with different $FF$ are covered in the same ring. The same applies to rings partially covering regions outside the experiment instrumented region.

The correction factor computed for proton-induced showers was also applied when considering gamma primaries, even if it introduces a small systematic error. Such error, which may be minimised by fine-tuning the correction factor, will most likely induce a marginal inclusion of a few gamma-induced showers in the upper tail of the corresponding gamma $LCm$ distribution. This will slightly increase the efficiency for gamma showers, which is irrelevant for the purposes of this article. In fact, for $K$ greater than a few units, the lines describing the evolution of $LCm$ as a function of $K$ of protons and gammas are essentially parallel (see Figure~5 from reference \cite{LCm}).

\begin{figure}[!t]
  \centering
  \includegraphics[width=0.45\textwidth]{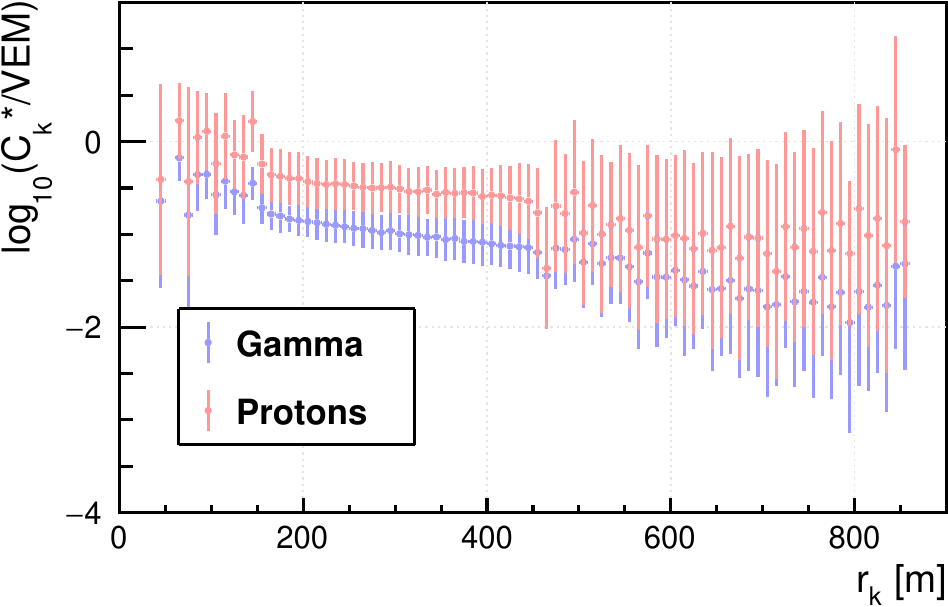}
  \caption{\label{fig:ckFFcorr} Same as Figure~\ref{fig:ckFFnocorr}, but with $C_{k}^{*}$ generalised following the definition in equation~\ref{eq:Ck*_cor}.
}
\end{figure}

Such corrections bring the mean values computed in each ring to the level of the values expected in the case of an equivalent ring embedded in a uniform array with the reference $FF$.
This effect is verified by comparing Figure~\ref{fig:ckFFnocorr} with Figure~\ref{fig:ckFFcorr}, produced in the same conditions but applying the correction factor.

In Figure~\ref{fig:ckFFcorr}, it can also be seen that while there is no relevant discontinuity in the mean values after applying the correction factor, the error bars are considerably smaller in the high $FF$ array region, as expected.

Lastly, it is noteworthy that while the aforementioned investigations were centered on simulations at approximately $100\,$TeV, it was verified that the findings remain consistent regardless of the energy. This verification was carried out by analyzing two additional energy bins, each containing $10\%$ of the simulation data used for the $100\,$TeV bin. These energy bins specifically encompassed $10\,$TeV and $1\,$PeV.

\section{Conclusions}
In this article, the applicability of the $C_k$ and $LCm$ gamma/hadron discriminator quantities to different realistic experimental scenarios has been addressed, namely:
\begin{itemize}
    \item It has been shown that azimuthal fluctuations of the shower footprint are better measured with water Cherenkov detectors. It can also be measured using scintillator arrays coupled to a converter but with less discrimination power than with WCDs. It was also verified that scintillator arrays alone have no gamma/hadron discrimination power. This is an indication that quantities like $C_k$ and $LCm$ are exploring the shower calorimetry information and not the number of particles at the ground.
    \item The station signal converted into the Vertical Equivalent Muon (VEM) units makes the computation of $C_k$ and $LCm$ essentially insensitive to the number of photosensors in the station.
    \item The realistic scenario in which the array presents a higher $FF$ close to its centre and is sparser in the external regions has also been examined. In this case, the appropriate generalisation necessary to correctly handle the $C_k$ variable has been derived.
\end{itemize}

The above statements allow us to conclude that shallow WCDs equipped with as few as one PMT should be considered a valid option to deal with the high energies in the design of future observatories such as Southern Wide-field Gamma-ray Observatory (SWGO) \cite{SWGO}.

\begin{acknowledgements}
The work here presented has been funded by OE - Portugal, FCT, I. P., under project PTDC/FIS-PAR/4300/2020. R.~C.\ is grateful for the financial support by OE - Portugal, FCT, I. P., under DL57/2016/cP1330/cT0002. L.~G. is grateful for the financial support by FCT PhD grant PRT/BD/154192/2022 under the IDPASC program. P.~J.~C. wants to acknowledge the financial support by FCT PhD under grant UI/BD/153576/2022.
\end{acknowledgements}

\bibliography{references.bib}   

\end{document}